\documentclass[12pt,preprint]{aastex}
\usepackage{amssymb}

\begin{document}

\begin{center}
\large\textbf{Compressible streaming instabilities in rotating thermal viscous
objects}
\large \\

\smallskip
\smallskip \textsf{A. K. Nekrasov}
\end{center}

Institute of Physics of the Earth, Russian Academy of Sciences, 123995
Moscow, Russia

\noindent e-mail: anatoli.nekrassov@t-online.de

\smallskip

\noindent \textbf{Abstract. }We study electromagnetic streaming
instabilities in thermal viscous regions of rotating astrophysical objects,
such as, protostellar and protoplanetary magnetized accretion disks,
molecular clouds, their cores, and elephant trunks. The obtained results can
also be applied to any regions of interstellar medium, where different
equilibrium velocities between charged species can arise. We consider a
weakly and highly ionized three-component plasma consisting of neutrals and
magnetized electrons and ions. The vertical perturbations along the
background magnetic field are investigated. The effect of perturbation of
collisional frequencies due to density perturbations of species is taken
into account. The growth rates of perturbations are found in a wide region
of wave number spectrum for media, where the thermal pressure is larger than
the magnetic pressure. It is shown that in cases of strong collisional
coupling of neutrals with ions the contribution of the viscosity is
negligible.

\noindent \textit{Subject headings: }accretion, accretion
disks-instabilities-magnetic fields-plasmas-waves

\section{Introduction}

In a series of papers, Nekrasov (2007, 2008 a,b, and 2009 a,b), a general
theory for electromagnetic compressible streaming instabilities in
multicomponent rotating magnetized objects, such as, accretion disks and
molecular clouds has been developed. In equilibrium of accretion disks,
different background velocities of different species (electrons, ions, dust
grains, and neutrals) have been found from the momentum equations with
taking into account anisotropic thermal pressure and collisions of charged
species with neutrals. Due to velocity differences, compressible streaming
instabilities have been shown to arise having growth rates much larger than
the rotation frequencies. New fast instabilities found in these papers have
been suggested to be a source of turbulence in accretion disks and molecular
clouds.

In papers cited above, the viscosity has not been considered. However,
numerical simulations of the magnetorotational instability show that this
effect can influence on the magnitude of the saturated amplitudes of
perturbations and, correspondingly, on the turbulent transport of the
angular momentum (e.g., Pessah \& Chan 2008; Masada \& Sano 2008). The
viscosity has numerically been shown (Yatou \& Toh 2009) can play a crucial
role in the persistence of the long-lived localized clouds observed in
interstellar media (Braun \& Kanekar 2005; Stanimirovi\'{c} \& Heiles 2005).
Interstellar media, molecular clouds, protostellar and protoplanetary
accretion disks are weakly ionized objects, where collisional effects play a
dominate role. The ratio of the viscosity to the resistivity (the magnetic
Prandtl number) for astrophysical objects takes a wide range of values. In
particular, in accretion disks around compact $X$-ray sources and active
galactic nuclei, the magnetic Prandtl number varies by several orders of
magnitude across the entire disk (Balbus \& Henry 2008). Therefore, the
viscosity is needed to be considered at studying electromagnetic streaming
instabilities in multicomponent weakly ionized media.

In the present paper, we explore electromagnetic streaming instabilities\ in
rotating astrophysical objects, such as protostellar and protoplanetary
magnetized accretion disks, molecular clouds, their cores, elephant trunks,
and so on, taking into account effects of collisions, thermal pressure and
viscosity. We consider a weakly and highly ionized three-component plasma
consisting of electrons, ions, and neutrals. The charged species are
supposed to be magnetized, i.e., their cyclotron frequencies are considered
to be much larger than their orbiting frequencies and collisional
frequencies with neutrals. We will investigate the vertical perturbations
along the background magnetic field. The presence of static (in
perturbations) dust grains can be invoked through the quasineutrality
condition. We take into account the effect of perturbation of collisional
frequencies due to density perturbations of species, which takes place at
different background velocities of species. We find expressions for the
perturbed velocity of any species that also contain the perturbed velocity
of other species due to collisions. For magnetized charged species, we
derive the dispersion relation, which is solved in the thermal regime when
the pressure force dominates the inertia. The conditions of strong or weak
collisional coupling of neutrals with charged species and the role of the
viscosity will be considered. The growth rates due to different azimuthal
velocities of electrons and ions will be found.

The paper is organized as follows. In Section 2 the basic equations are
given. In Section 3 we shortly discuss the equilibrium state. Solutions for
the perturbed velocities of species for the vertical perturbations are
obtained in Section 4. The dispersion relation in the general form is
derived in Section 5. In Section 6 this dispersion relation is solved in the
thermal regime in the specific cases and unstable solutions are found. In
Section 7 we give an expression needed for determining of polarization of
perturbations. Some problems concerning the contribution of the resistivity
in the standard MHD are discussed in Section 8. Discussion of the
obtained results and their applicability to protostellar and protoplanetary
disks are given in Section 9. The main points of the paper are summarized in
Section 10.

\section{Basic equations}

We will consider weakly ionized rotating objects consisting of electrons,
ions, and neutrals. Here, we do not treat the presence of dust grains.
However, the latter may be involved as static (in perturbations) species
through the condition of quasineutrality. The electrons and ions are
supposed to be magnetized, i.e., their cyclotron frequencies are larger than
their collisional frequencies with neutrals. Self-gravity is not included.
We will study one-dimensional perturbations along the background magnetic
field $\mathbf{B}_{0}$. Then the momentum equations for species in the
inertial (nonrotating) reference frame accounting for the viscosity
(Braginskii 1965) take the form, 
\begin{eqnarray}
\frac{\partial \mathbf{v}_{j}}{\partial t}+\mathbf{v}_{j}\cdot \mathbf{%
\nabla v}_{j} &=&-\mathbf{\nabla }U-\frac{\mathbf{\nabla }P_{j}}{m_{j}n_{j}}+%
\frac{q_{j}}{m_{j}}\mathbf{E+}\frac{q_{j}}{m_{j}c}\mathbf{v}_{j}\times 
\mathbf{B-}\nu _{jn}\left( \mathbf{v}_{j}-\mathbf{v}_{n}\right) \\
&&+\frac{\mu _{j}}{\omega _{cj}\tau _{jn}}\frac{\partial ^{2}\mathbf{v}%
_{j}\times \mathbf{b}}{\partial z^{2}}+\frac{6}{5}\frac{\mu _{j}}{\omega
_{cj}^{2}\tau _{jn}^{2}}\frac{\partial ^{2}\mathbf{v}_{j\perp }}{\partial
z^{2}}+\frac{4}{3}\mu _{j}\frac{\partial ^{2}\mathbf{v}_{jz}}{\partial z^{2}}%
,  \nonumber
\end{eqnarray}%
\begin{equation}
\frac{\partial \mathbf{v}_{n}}{\partial t}+\mathbf{v}_{n}\cdot \mathbf{%
\nabla v}_{n}=-\mathbf{\nabla }U-\frac{\mathbf{\nabla }P_{n}}{m_{n}n_{n}}%
-\sum_{j}\nu _{nj}\left( \mathbf{v}_{n}-\mathbf{v}_{j}\right) +\mu
_{n}\left( \frac{\partial ^{2}\mathbf{v}_{n}}{\partial z^{2}}+\frac{1}{3}%
\frac{\partial ^{2}\mathbf{v}_{nz}}{\partial z^{2}}\right) ,
\end{equation}%
\noindent where the index $j=e,i$ denotes the electrons and ions,
respectively, and the index $n$ denotes the neutrals. In Equations (1) and
(2) $q_{j}$ and $m_{j,n}$ are the charge and mass of species $j$ and
neutrals, $\mathbf{v}_{j,n}$ is the hydrodynamic velocity, $n_{j,n}$ is the
number density, $P_{j,n}=n_{j,n}T_{j,n}$ is the thermal pressure, $T_{j,n}$
is the temperature, $\nu _{jn}$ $=\gamma _{jn}m_{n}n_{n}$ ($\nu _{nj}=\gamma
_{jn}m_{j}n_{j}$) is the collisional frequency of species $j$ (neutrals)
with neutrals (species $j$). The indices $\perp $ and $z$ denote directions
across and along the magnetic field, respectively. The value $\gamma _{jn}$
is $\gamma _{jn}=<\sigma v>_{jn}/(m_{j}+m_{n})$, where $<\sigma v>_{jn}$is
the rate coefficient for momentum transfer, and $\mu _{j,n}=v_{Tj,n}^{2}/\nu
_{jn,nn}$ are the coefficients of the kinematic viscosity ($\nu _{nn}$ is
the neutral-neutral collisional frequency), where $%
v_{Tj,n}=(T_{j,n}/m_{j,n})^{1/2}$ is the thermal velocity. Further, $\omega
_{cj}=q_{j}B_{0}/m_{j}c$ is the cyclotron frequency and $\tau _{jn}=\nu
_{jn}^{-1}$. Other notations are the following: $U=-GM/R$\ is the
gravitational potential of the central object having mass $M$ (when it
presents), $G$ is the gravitational constant, $R=(r^{2}+z^{2})^{1/2}$, $%
\mathbf{E}$\textbf{\ }and $\mathbf{B}$ are the electric and magnetic fields, 
$\mathbf{b}=\mathbf{B}/B$, and $c$ is the speed of light in vacuum. The
magnetic field $\mathbf{B}$ includes the external magnetic field $\mathbf{B}%
_{0ext}$ of the central object and/or interstellar medium, the magnetic
field $\mathbf{B}_{0cur}$ of the stationary current in a steady state, and
the perturbed magnetic field. We use the cylindrical coordinate system $%
(r,\theta ,z),$ where $r$ is the distance from the symmetry axis $z$, and $%
\theta $ is the azimuthal direction. We assume that the background magnetic
field is directed along the $z$ axis, $\mathbf{B}_{0}=\mathbf{B}_{0zext}+%
\mathbf{B}_{0zcur}$. In Equation (1), the condition $\omega _{cj}\gg \nu
_{jn}$ is satisfied for the viscous terms. For unmagnetized charged
particles of species $j,$ $\omega _{cj}\ll \nu _{jn},$ the viscosity
coefficient has the same form as that for neutrals. We adopt the adiabatic
model for the temperature evolution when $P_{j,n}\sim n_{j,n}^{\gamma _{a}}$%
, where $\gamma _{a}$ is the adiabatic constant.

The other basic equations are the continuity equation, 
\begin{equation}
\frac{\partial n_{j,n}}{\partial t}+\mathbf{\nabla \cdot }n_{j,n}\mathbf{v}%
_{j,n}=0,
\end{equation}

\noindent Faraday's law, 
\begin{equation}
\mathbf{\nabla \times E=-}\frac{1}{c}\frac{\partial \mathbf{B}}{\partial t},
\end{equation}%
and Ampere's law,%
\begin{equation}
\mathbf{\nabla \times B=}\frac{4\pi }{c}\mathbf{j,}
\end{equation}

\noindent where $\mathbf{j=}\sum_{j}q_{j}n_{j}\mathbf{v}_{j}.$ We consider the wave
processes with typical time scales much larger than the time the light
spends to cover the wavelength of perturbations. In this case, one can
neglect the displacement current in Equation (5) that results in
quasineutrality both for the electromagnetic and purely electrostatic
perturbations.

\section{Equilibrium}

We suppose that electrons, ions, and neutrals rotate in the azimuthal
direction of the astrophysical object (accretion disk, molecular cloud, its
cores, elephant trunk, and so on) with different, in general, velocities $%
v_{j,n0}.$ The stationary dynamics of light charged species, electrons and
ions, is undergone by the effect of the background magnetic field and
collisions with neutrals. In their turn, the neutrals also experience the
collisional coupling with charged species influencing on their equilibrium
velocity. Some specific cases of equilibrium have been investigated in
papers by Nekrasov (2007, 2008 a,b, 2009 b), where the expressions for
stationary velocities of species in the gravitational field of the central
mass have been found at the absence of collisions as well as with taking
into account collisions for cases of weak and strong collisional coupling of
neutrals with ions.

Due to different stationary velocities of charged species, the electric
currents are generated in the equilibrium state.

\section{Linear regime}

In the present paper, we do not treat perturbations connected with the
background pressure gradients. Thus, we exclude the drift and internal
gravity waves from our consideration. We take into account the induced
reaction of neutrals on the perturbed motion of charged species. The
neutrals can be involved in electromagnetic perturbations, if the ionization
degree of medium is sufficiently high. We also include the effect of
perturbation of the collisional frequencies due to density perturbations of
charged species and neutrals. This effect emerges when there are different
background velocities of species. Then the momentum equations (1) and (2) in
the linear approximation take the form, 
\begin{eqnarray}
\frac{\partial \mathbf{v}_{j1}}{\partial t} &=&-c_{sj}^{2}\frac{\mathbf{%
\nabla }n_{j1}}{n_{j0}}+\frac{q_{j}}{m_{j}}\mathbf{E}_{1}\mathbf{+}\frac{%
q_{j}}{m_{j}c}\mathbf{v}_{j0}\times \mathbf{B}_{1}\mathbf{+}\frac{q_{j}}{%
m_{j}c}\mathbf{v}_{j1}\mathbf{\times \mathbf{B}}_{0}\mathbf{-}\nu
_{jn}^{0}\left( \mathbf{v}_{j1}-\mathbf{v}_{n1}\right) \\
&&\mathbf{-}\frac{n_{n1}}{n_{n0}}\nu _{jn}^{0*}\left( \mathbf{v}_{j0}-%
\mathbf{v}_{n0}\right) +\frac{\mu _{j}}{\omega _{cj}\tau _{jn}}\frac{%
\partial ^{2}\mathbf{v}_{j1}\times \mathbf{b}_{0}}{\partial z^{2}}+\frac{6}{5%
}\frac{\mu _{j}}{\omega _{cj}^{2}\tau _{jn}^{2}}\frac{\partial ^{2}\mathbf{v}%
_{j\perp 1}}{\partial z^{2}}+\frac{4}{3}\mu _{j}\frac{\partial ^{2}\mathbf{v}%
_{jz1}}{\partial z^{2}},  \nonumber
\end{eqnarray}
\begin{eqnarray}
\frac{\partial \mathbf{v}_{n1}}{\partial t} &=&-c_{sn}^{2}\frac{\mathbf{%
\nabla }n_{n1}}{n_{n0}}-\sum_{j}\nu _{nj}^{0}\left( \mathbf{v}_{n1}-\mathbf{v%
}_{j1}\right) -\sum_{j}\frac{n_{j1}}{n_{j0}}\nu _{nj}^{0*}\left( \mathbf{v}%
_{n0}-\mathbf{v}_{j0}\right) \\
&&+\mu _{n}\left( \frac{\partial ^{2}\mathbf{v}_{n1}}{\partial z^{2}}+\frac{1%
}{3}\frac{\partial ^{2}\mathbf{v}_{nz1}}{\partial z^{2}}\right) ,  \nonumber
\end{eqnarray}

\noindent where $c_{sj,n}=(\gamma _{a}T_{j,n0}/m_{j,n})^{1/2}$ is the sound
velocity ($\gamma _{a}=3$ in the one-dimensional case), $\nu
_{jn}^{0}=\gamma _{jn}m_{n}n_{n0}$, $\nu _{nj}^{0}=\gamma _{jn}m_{j}n_{j0}$.
The terms proportional to $\nu _{jn}^{0\ast }$ and $\nu _{nj}^{0\ast }$
describe the effect of perturbation of the collisional frequencies due to
density perturbations. The index $1$ denotes the quantities of the first
order of magnitude. The neutrals participate in the electromagnetic dynamics
due to collisional coupling with the charged species, mostly with ions (in
the absence of dust grains). However, we keep, for generality, collisions of
neutrals with electrons in Equation (7).

The continuity Equation (3) in the linear regime is the following: 
\begin{equation}
\frac{\partial n_{j,n1}}{\partial t}+n_{j,n0}\frac{\partial v_{j,n1z}}{%
\partial z}=0.
\end{equation}

We further apply the Fourier transform to Equations (6)-(8), supposing the
perturbations of the form $\exp (ik_{z}z-i\omega t).$ Using Equation (8) and
solving Equation (7), we find the expressions for components of the Fourier
amplitude $\mathbf{v}_{n1k}$, where (and below) the index $k=\{k_{z},\omega
\}$, 
\begin{eqnarray}
-i\omega _{n\perp }v_{n1rk} &=&\sum_{j}\nu _{nj}v_{j1rk}, \\
-i\omega _{n\perp }v_{n1\theta k} &=&\sum_{j}\nu _{nj}v_{j1\theta
k}-\sum_{j}\nu _{nj}^{*}\frac{k_{z}\left( v_{n0}-v_{j0}\right) }{\omega }%
v_{j1zk},  \nonumber \\
-i\omega _{nz}v_{n1zk} &=&\sum_{j}\nu _{nj}v_{j1zk}.  \nonumber
\end{eqnarray}

\noindent Here and below, the index $0$ by $\nu _{nj}^{0}$ and $\nu
_{jn}^{0} $ is, for simplicity, omitted. In Equations (9), the notations are
introduced, 
\begin{eqnarray*}
\omega _{n\perp } &=&\omega +i\nu _{n}+i\frac{3}{4}\chi _{nz}, \\
\omega _{nz} &=&\omega -\frac{k_{z}^{2}c_{sn}^{2}}{\omega }+i\nu _{n}+i\chi
_{nz},
\end{eqnarray*}

\noindent where $\nu _{n}=\sum_{j}\nu _{nj}$, $\chi _{n,jz}=\frac{4}{3}\mu
_{n,j}k_{z}^{2}$.

Now we substitute the expressions for $\mathbf{v}_{n1k}$ and $%
n_{j,n1k}=n_{j,n0}k_{z}v_{j,n1zk}/\omega $ in Equation (6). Then we obtain
the following expressions for components of $\mathbf{v}_{j1k}$: 
\begin{eqnarray}
-i\omega _{j\perp }v_{j1rk} &=&\frac{q_{j}}{m_{j}}E_{1rk}+\omega
_{j}v_{j1\theta k}+i\frac{\nu _{jn}}{\omega _{n\perp }}\sum_{l}\nu
_{nl}v_{l1rk}, \\
-i\omega _{j\perp }v_{j1\theta k} &=&\frac{q_{j}}{m_{j}}E_{1\theta k}-\omega
_{j}v_{j1rk}+i\frac{\nu _{jn}}{\omega _{n\perp }}\sum_{l}\nu
_{nl}v_{l1\theta k}-i\sum_{l}\beta _{jlz}v_{l1zk},  \nonumber \\
-i\omega _{jz}v_{j1zk} &=&\frac{q_{j}}{m_{j}}\left( E_{1zk}+n_{z}\frac{v_{j0}%
}{c}E_{1\theta k}\right) +i\frac{\nu _{jn}}{\omega _{nz}}\sum_{l}\nu
_{nl}v_{l1zk}.  \nonumber
\end{eqnarray}

\noindent Here the notations are introduced: 
\begin{eqnarray*}
\omega _{j\perp } &=&\omega +i\nu _{jn}+i\chi _{j\perp 2}, \\
\omega _{jz} &=&\omega -\frac{k_{z}^{2}c_{sj}^{2}}{\omega }+i\nu _{jn}+i\chi
_{jz}, \\
\omega _{j} &=&\omega _{cj}-\chi _{j\perp 1}, \\
\beta _{jlz} &=&\nu _{jn}\nu _{nl}^{*}\frac{k_{z}}{\omega }\left( \frac{%
v_{n0}-v_{l0}}{\omega _{n\perp }}+\frac{v_{j0}-v_{n0}}{\omega _{nz}}\right) ,
\end{eqnarray*}

\noindent where $\chi _{j\perp 1}=\mu _{j}k_{z}^{2}/\omega _{cj}\tau _{jn},$ 
$\chi _{j\perp 2}=1.2\mu _{j}k_{z}^{2}/\omega _{cj}^{2}\tau _{jn}^{2}$. In
the last Equation (10), we have substituted $B_{1rk}$ for $E_{1\theta k}$,
using Faraday's equation (4), $B_{1rk}=-n_{z}E_{1\theta k}$ ($B_{1\theta
k}=n_{z}E_{1rk}$, $B_{1zk}=0$), where $n_{z}=k_{z}c/\omega $.

From Equation (10) for $v_{j1zk}$, we can find the longitudinal (along $%
B_{0} $) velocities of electrons and ions, 
\begin{equation}
D_{z}v_{j1zk}=in_{z}b_{jz}E_{1\theta k}+ia_{jz}E_{1zk}.
\end{equation}

\noindent Here, 
\begin{eqnarray*}
D_{z} &=&\omega _{ez}\omega _{iz}\omega _{nz}+\omega _{iz}\nu _{en}\nu
_{ne}+\omega _{ez}\nu _{in}\nu _{ni}, \\
a_{ez} &=&\alpha _{iz}\frac{q_{e}}{m_{e}}-\nu _{en}\nu _{ni}\frac{q_{i}}{%
m_{i}}, \\
b_{ez} &=&\alpha _{iz}\frac{q_{e}}{m_{e}}\frac{v_{e0}}{c}-\nu _{en}\nu _{ni}%
\frac{q_{i}}{m_{i}}\frac{v_{i0}}{c}, \\
a_{iz} &=&\alpha _{ez}\frac{q_{i}}{m_{i}}-\nu _{in}\nu _{ne}\frac{q_{e}}{%
m_{e}}, \\
b_{iz} &=&\alpha _{ez}\frac{q_{i}}{m_{i}}\frac{v_{i0}}{c}-\nu _{in}\nu _{ne}%
\frac{q_{e}}{m_{e}}\frac{v_{e0}}{c}, \\
\alpha _{jz} &=&\omega _{jz}\omega _{nz}+\nu _{jn}\nu _{nj}.
\end{eqnarray*}

Let us find now the transverse (across $B_{0}$) velocities of charged
species. When solving Equations (10), we will consider the case in which the
charged species are magnetized, i.e., we will suppose that the following
conditions are satisfied: 
\begin{equation}
\omega _{cj}\gg \omega _{j\perp },\nu _{jn}\nu _{n}/\omega _{n\perp },\chi
_{j\perp 1}.
\end{equation}

\noindent Conditions (12) signify that the Lorentz force is dominant. Then,
using Equation (11), we find solutions, 
\begin{eqnarray}
v_{j1rk} &=&-i\frac{q_{j}}{m_{j}\omega _{cj}^{2}}b_{j}E_{1rk}+\frac{1}{%
\omega _{cj}}\left( \frac{q_{j}}{m_{j}}a_{j}+\frac{n_{z}}{D_{z}}\lambda
_{jz}\right) E_{1\theta k}+\frac{\delta _{jz}}{\omega _{cj}D_{z}}E_{1zk}, \\
v_{j1\theta k} &=&-\frac{q_{j}}{m_{j}\omega _{cj}}a_{j}E_{1rk}-i\frac{q_{j}}{%
m_{j}\omega _{cj}^{2}}b_{j}E_{1\theta k}.  \nonumber
\end{eqnarray}

\noindent Here, 
\begin{eqnarray*}
a_{j} &=&1+\frac{\chi _{j\perp 1}}{\omega _{cj}},b_{j}=\omega _{j\perp }+%
\frac{\nu _{jn}\nu _{n}}{\omega _{n\perp }}, \\
\delta _{jz} &=&\frac{q_{e}}{m_{e}}\gamma _{jez}+\frac{q_{i}}{m_{i}}\gamma
_{jiz}, \\
\lambda _{jz} &=&\frac{q_{e}}{m_{e}}\gamma _{jez}\frac{v_{e0}}{c}+\frac{q_{i}%
}{m_{i}}\gamma _{jiz}\frac{v_{i0}}{c}, \\
\gamma _{jez} &=&\beta _{jez}\alpha _{iz}-\beta _{jiz}\nu _{in}\nu _{ne}, \\
\gamma _{jiz} &=&\beta _{jiz}\alpha _{ez}-\beta _{jez}\nu _{en}\nu _{ni}.
\end{eqnarray*}

\noindent In the next section, we calculate the perturbed electric current
and obtain the dispersion relation.

\section{Dispersion relation}

From Equations (4) and (5) we obtain, 
\begin{eqnarray}
n_{z}^{2}\mathbf{E}_{1\perp k} &=&\frac{4\pi i}{\omega }\mathbf{j}_{1\perp
k}, \\
j_{1zk} &=&0.  \nonumber
\end{eqnarray}

\noindent Using solutions (11) and (13), we can calculate the electric
current $\mathbf{j}_{1k}=\sum_{j}q_{j}n_{j0}\mathbf{v}_{j1k}+%
\sum_{j}q_{j}n_{j1k}\mathbf{v}_{j0}$. Substituting this current in Equations
(14), we will find the following set of equations for determining the
components of the perturbed electric field $\mathbf{E}_{1k}=$ $(E_{1\theta
k},E_{1rk},E_{1zk})$: 
\begin{equation}
\mathbf{\hat{A}}_{k}\mathbf{E}_{1k}=\mathbf{0},
\end{equation}

\noindent where the matrix $\mathbf{\hat{A}}_{k}$ is equal to 
\[
\mathbf{\hat{A}}_{k}=\left| 
\begin{array}{ccc}
n_{z}^{2}-\varepsilon _{1}+\varepsilon _{5}, & i\left( \varepsilon
_{2}+\varepsilon _{\chi \perp }\right) , & \varepsilon _{6} \\ 
-i\left( \varepsilon _{2}+\varepsilon _{\chi \perp }+\varepsilon _{\lambda
z}\right) , & n_{z}^{2}-\varepsilon _{1}, & -i\varepsilon _{\delta z} \\ 
\varepsilon _{4}, & 0 & -\varepsilon _{3}%
\end{array}
\right| . 
\]

\noindent The components of the matrix $\mathbf{\hat{A}}_{k}$ are the
following:

\begin{eqnarray*}
\varepsilon _{1} &=&\sum_{j}\frac{\omega _{pj}^{2}}{\omega \omega _{cj}^{2}}%
b_{j},\varepsilon _{2}=-\frac{\omega _{pd}^{2}}{\omega \omega _{cd}}%
,\varepsilon _{3}=-\frac{1}{\omega D_{z}}\sum_{j}\omega _{pj}^{2}\frac{m_{j}%
}{q_{j}}a_{jz},\varepsilon _{4}=\frac{n_{z}}{\omega D_{z}}\sum_{j}\omega
_{pj}^{2}\frac{m_{j}}{q_{j}}b_{jz}, \\
\varepsilon _{5} &=&\frac{n_{z}^{2}}{\omega D_{z}}\sum_{j}\omega _{pj}^{2}%
\frac{m_{j}}{q_{j}}\frac{v_{j0}}{c}b_{jz},\varepsilon _{6}=\frac{n_{z}}{%
\omega D_{z}}\sum_{j}\omega _{pj}^{2}\frac{m_{j}}{q_{j}}\frac{v_{j0}}{c}%
a_{jz}, \\
\varepsilon _{\chi \perp } &=&\sum_{j}\frac{\omega _{pj}^{2}}{\omega \omega
_{cj}^{2}}\chi _{j\perp 1},\varepsilon _{\delta z}=\frac{1}{\omega D_{z}}%
\sum_{j}\frac{\omega _{pj}^{2}}{\omega _{cj}}\frac{m_{j}}{q_{j}}\delta
_{jz},\varepsilon _{\lambda z}=\frac{n_{z}}{\omega D_{z}}\sum_{j}\frac{%
\omega _{pj}^{2}}{\omega _{cj}}\frac{m_{j}}{q_{j}}\lambda _{jz},
\end{eqnarray*}

\noindent where $\omega _{pj}=\left( 4\pi n_{j0}q_{j}^{2}/m_{j}\right)
^{1/2} $ is the plasma frequency and the index $d$ denotes dust grains (if
they present and are static in perturbations).

The dispersion relation is obtained by equating the determinant of the
matrix $\mathbf{\hat{A}}_{k}$ to zero. As a result, we obtain, 
\begin{equation}
\left[ \left( n_{z}^{2}-\varepsilon _{1}\right) \varepsilon _{3}+\varepsilon
_{3}\varepsilon _{5}+\varepsilon _{4}\varepsilon _{6}\right] \left(
n_{z}^{2}-\varepsilon _{1}\right) -\left( \varepsilon _{2}+\varepsilon
_{\chi \perp }\right) \left[ \left( \varepsilon _{2}+\varepsilon _{\chi
\perp }\right) \varepsilon _{3}+\varepsilon _{3}\varepsilon _{\lambda
z}+\varepsilon _{4}\varepsilon _{\delta z}\right] =0.
\end{equation}

\noindent In this equation, it is easy to see that the value $\varepsilon
_{2}+\varepsilon _{\chi \perp }$ can be, in general, small in comparison to
the value $n_{z}^{2}-\varepsilon _{1}$. Therefore, we consider below the
case in which the following conditions are satisfied: 
\begin{eqnarray}
\left( n_{z}^{2}-\varepsilon _{1}\right) &\gg &\left( \varepsilon
_{2}+\varepsilon _{\chi \perp }\right) , \\
\left( \varepsilon _{3}\varepsilon _{5}+\varepsilon _{4}\varepsilon
_{6}\right) \left( n_{z}^{2}-\varepsilon _{1}\right) &\gg &\left(
\varepsilon _{2}+\varepsilon _{\chi \perp }\right) \left( \varepsilon
_{3}\varepsilon _{\lambda z}+\varepsilon _{4}\varepsilon _{\delta z}\right) .
\nonumber
\end{eqnarray}

\noindent Then Equation (16) will take the form, 
\begin{equation}
\left( n_{z}^{2}-\varepsilon _{1}\right) \varepsilon _{3}+\varepsilon
_{3}\varepsilon _{5}+\varepsilon _{4}\varepsilon _{6}=0.
\end{equation}

\noindent We do not consider the damping Alfv\'{e}n perturbations $%
n_{z}^{2}-\varepsilon _{1}=0$. When Equation (18) is satisfied, the
component of the electric field $E_{1\theta ,zk}\neq 0$ and $E_{1rk}\ll
E_{1\theta k}$. The second and third equations of a set (15) determine the
components $E_{1rk}$ and $E_{1zk}$, 
\begin{eqnarray}
\left( n_{z}^{2}-\varepsilon _{1}\right) \varepsilon _{3}E_{1rk} &=&i\left[
\left( \varepsilon _{2}+\varepsilon _{\chi \perp }\right) \varepsilon
_{3}+\varepsilon _{3}\varepsilon _{\lambda z}+\varepsilon _{4}\varepsilon
_{\delta z}\right] E_{1\theta k}, \\
\varepsilon _{3}E_{1zk} &=&\varepsilon _{4}E_{1\theta k}.  \nonumber
\end{eqnarray}

\noindent Below, we will investigate solutions of Equation (18).

\section{6. Dispersion relation (18) and its solutions in the specific
case}

We further will not take into account the presence of dust grains. Then the
quasineutrality condition has the form $q_{e}n_{e0}+q_{i}n_{i0}=0$. Using
this condition, we will calculate the quantities $\varepsilon _{1}$, $%
\varepsilon _{3}$, $\varepsilon _{4}$, $\varepsilon _{5}$, and $\varepsilon
_{6}$. We will be interested in the case in which $\omega \ll \nu _{n}+\chi
_{nz}$. When $\chi _{nz}$ $\leq \nu _{n}$ there is strong collisional
coupling of neutrals with charged species in the transverse direction to the 
$z$ axis. This coupling along the magnetic field depends on the relation
between $k_{z}^{2}c_{sn}^{2}$ and $\omega \nu _{n}$ [see Equations (9)]. If $%
\chi _{nz}$ $\gg \nu _{n}$, then the neutrals have a weak coupling with
charged species. We suppose also that $\omega \ll \nu _{jn},\nu _{nn}$ (this
inequality is wittingly satisfied in weakly ionized plasma if $\chi _{nz}$ $%
\leq \nu _{n}$ and $\omega \ll \nu _{n}$). In this case, the contribution of
the viscosity $\chi _{j,nz}$ in the expressions for $\omega _{j,nz}$ is much
smaller in comparison to the thermal term $k_{z}^{2}c_{sj,n}^{2}/\omega $
and can be ignored. We further will investigate the case in which $%
k_{z}^{2}c_{sj,n}^{2}\gg \omega ^{2}$ when the thermal pressure dominates
the inertia. Below, we will obtain the dispersion relations for two regions
of small and large wave number $k_{z}$.

\subsection{Dispersion relation (18) for small $k_{z}$}

At first, suppose that the following condition is satisfied: 
\begin{equation}
\omega d_{j}\gg k_{z}^{2}c_{sj}^{2}c_{sn}^{2},
\end{equation}

\noindent where 
\[
d_{j}=c_{sj}^{2}\nu _{n}+c_{sn}^{2}\nu _{jn}. 
\]

\noindent Under conditions at hand we will find, 
\[
\varepsilon _{1}=\sum_{j}\frac{\omega _{pj}^{2}}{\omega _{cj}^{2}}\frac{\nu
_{jn}}{\nu _{n}}=\frac{c^{2}}{c_{A}^{2}}, 
\]
\[
\varepsilon _{3}=i\frac{\omega _{pe}^{2}k_{z}^{2}}{\omega ^{2}D_{z}}\left(
d_{i}-\frac{q_{i}m_{e}}{q_{e}m_{i}}d_{e}\right) , 
\]
\[
\varepsilon _{4}=-i\frac{\omega _{pe}^{2}n_{z}k_{z}^{2}}{\omega ^{2}D_{z}}%
d_{1}, 
\]
\[
\varepsilon _{5}=-i\frac{\omega _{pe}^{2}n_{z}^{2}}{\omega D_{z}}\frac{%
k_{z}^{2}}{\omega }d_{2}+\frac{\omega _{pe}^{2}n_{z}^{2}}{\omega D_{z}}%
u_{\nu 2}\frac{\left( v_{i0}-v_{e0}\right) }{c}, 
\]

\begin{equation}
\varepsilon _{6}=\varepsilon _{4}+\frac{\omega _{pe}^{2}n_{z}}{\omega D_{z}}%
u_{\nu 1}\frac{\left( v_{i0}-v_{e0}\right) }{c},
\end{equation}

\noindent where

\[
d_{1}=\frac{v_{e0}}{c}d_{i}-\frac{q_{i}m_{e}}{q_{e}m_{i}}\frac{v_{i0}}{c}%
d_{e},d_{2}=\frac{v_{e0}^{2}}{c^{2}}d_{i}-\frac{q_{i}m_{e}}{q_{e}m_{i}}\frac{%
v_{i0}^{2}}{c^{2}}d_{e}, 
\]

\begin{eqnarray*}
u_{\nu 1} &=&\nu _{in}\nu _{ne}+\frac{q_{i}m_{e}}{q_{e}m_{i}}\nu _{en}\nu
_{ni}, \\
u_{\nu 2} &=&\nu _{in}\nu _{ne}\frac{v_{e0}}{c}+\frac{q_{i}m_{e}}{q_{e}m_{i}}%
\nu _{en}\nu _{ni}\frac{v_{i0}}{c}.
\end{eqnarray*}

\noindent In the expression for $\varepsilon _{1}$, we have supposed that $%
\nu _{n}\gg $ $\chi _{nz}$ that is compatible with condition (20).

The value $D_{z}$ has the form, 
\begin{equation}
D_{z}=g\frac{k_{z}^{2}}{\omega },
\end{equation}

\noindent where 
\[
g=c_{se}^{2}\nu _{in}\nu _{ne}+c_{si}^{2}\nu _{en}\nu _{ni}+c_{sn}^{2}\nu
_{in}\nu _{en}. 
\]

Substitute now the expressions (21) and (22) in Equation (18). Then we
obtain the following dispersion relation under conditions given above: 
\begin{equation}
\left( n_{z}^{2}-\varepsilon _{1}\right) \left( \frac{q_{e}}{m_{e}}d_{i}-%
\frac{q_{i}}{m_{i}}d_{e}\right) g+\frac{\omega _{pe}^{2}}{\omega ^{3}}\frac{%
q_{i}}{m_{i}}\left[ ik_{z}^{2}d_{e}d_{i}+\omega \left( \nu _{en}\nu
_{ni}d_{i}+\nu _{in}\nu _{ne}d_{e}\right) \right] \left(
v_{i0}-v_{e0}\right) ^{2}=0.
\end{equation}

\paragraph{6.1.1. Solution of dispersion relation (23) for sufficiently
large $k_{z}$}

Let us find a solution of Equation (23) in the case 
\begin{equation}
k_{z}^{2}d_{e}d_{i}\gg \omega \left( \nu _{en}\nu _{ni}d_{i}+\nu _{in}\nu
_{ne}d_{e}\right) .
\end{equation}

\noindent Then Equation (23) takes the form 
\begin{equation}
\omega ^{3}-\omega k_{z}^{2}c_{A}^{2}+ih^{2}k_{z}^{2}c_{A}^{2}\frac{\left(
v_{e0}-v_{i0}\right) ^{2}}{c^{2}}=0,
\end{equation}

\noindent where

\[
h^{2}=\omega _{pe}^{2}\frac{q_{i}}{m_{i}}\frac{d_{e}d_{i}}{g\left( %
\frac{|q_{e}|}{m_{e}}d_{i}+\frac{q_{i}}{m_{i}}d_{e}\right) }. 
\]

\noindent The sign $|$ $|$ denotes an absolute value. \noindent We see that
due to the last term on the left-hand side of Equation (25) there is an
unstable solution.

The solution of Equation (25) in the region $\omega ^{2}\ll
k_{z}^{2}c_{A}^{2}$ is equal to 
\begin{equation}
\gamma =h^{2}\frac{\left( v_{e0}-v_{i0}\right) ^{2}}{c^{2}},
\end{equation}

\noindent where $\gamma =$Im $\omega $ is the growth rate. \noindent Thus,
for $k_{z}$ such that $k_{z}>k_{z}^{*}=h^{2}\left( v_{e0}-v_{i0}\right)
^{2}/c_{A}c^{2}$, the growth rate is maximal and independent from the wave
number. \noindent In the region $\omega ^{2}\gg k_{z}^{2}c_{A}^{2}$, we
obtain,

\begin{equation}
\gamma =\left[ hk_{z}c_{A}\frac{\left( v_{e0}-v_{i0}\right) }{c}\right]
^{2/3}.
\end{equation}

\noindent This growth rate increases with increasing of $k_{z}$ in the
region $k_{z}<k_{z}^{*}$.

\paragraph{6.1.2. Solution of dispersion relation (23) for sufficiently
small $k_{z}$}

In the case 
\begin{equation}
k_{z}^{2}d_{e}d_{i}\ll \omega \left( \nu _{en}\nu _{ni}d_{i}+\nu _{in}\nu
_{ne}d_{e}\right) ,
\end{equation}

\noindent Equation (23) has a solution, 
\begin{equation}
\omega =\left[ k_{z}^{2}-s^{2}\frac{\left( v_{e0}-v_{i0}\right) ^{2}}{c^{2}}%
\right] ^{1/2}c_{A},
\end{equation}

\noindent where

\[
s^{2}=\omega _{pe}^{2}\frac{q_{i}}{m_{i}}\frac{\left( \nu _{en}\nu
_{ni}d_{i}+\nu _{in}\nu _{ne}d_{e}\right) }{\left( \frac{|q_{e}|}{m_{e}}d_{i}%
+\frac{q_{i}}{m_{i}}d_{e}\right) g}. 
\]

\noindent We see that perturbations with the wave number $k_{z}<
s|v_{e0}-v_{i0}|/c$ will be unstable.

\subsection{Solutions of dispersion relation (18) for large $k_{z}$}

Now we consider the dispersion relation (18) in the case of large $k_{z}$
when 
\begin{equation}
k_{z}^{2}c_{sj}^{2}c_{sn}^{2}\gg \omega d_{j}.
\end{equation}

\noindent Then we obtain the following expressions: 
\[
\varepsilon _{3}=-\frac{\omega _{pe}^{2}}{\omega ^{3}D_{z}}%
k_{z}^{4}c_{sn}^{2}\left( c_{si}^{2}-\frac{q_{i}m_{e}}{q_{e}m_{i}}%
c_{se}^{2}\right) , 
\]%
\[
\varepsilon _{4}=\frac{\omega _{pe}^{2}n_{z}}{\omega ^{3}D_{z}}%
k_{z}^{4}c_{sn}^{2}\left( \frac{v_{e0}}{c}c_{si}^{2}-\frac{q_{i}m_{e}}{%
q_{e}m_{i}}\frac{v_{i0}}{c}c_{se}^{2}\right) , 
\]%
\[
\varepsilon _{5}=\frac{\omega _{pe}^{2}n_{z}^{2}}{\omega ^{3}D_{z}}%
k_{z}^{4}c_{sn}^{2}\left( \frac{v_{e0}^{2}}{c^{2}}c_{si}^{2}-\frac{q_{i}m_{e}%
}{q_{e}m_{i}}\frac{v_{i0}^{2}}{c^{2}}c_{se}^{2}\right) +\frac{\omega
_{pe}^{2}n_{z}^{2}}{\omega D_{z}}u_{\nu 2}\frac{v_{i0}-v_{e0}}{c}, 
\]%
\[
\varepsilon _{6}=\varepsilon _{4}+\frac{\omega _{pe}^{2}n_{z}}{\omega D_{z}}%
u_{\nu 1}\frac{v_{i0}-v_{e0}}{c}, 
\]%
\begin{equation}
D_{z}=-\frac{k_{z}^{6}}{\omega ^{3}}c_{se}^{2}c_{si}^{2}c_{sn}^{2}.
\end{equation}

\noindent The expression for $\varepsilon _{1}$ in the case $\nu _{n}\gg
\chi _{nz}$ has the form (21). In the opposite case, $\nu _{n}\ll \chi _{nz}$%
, we obtain, 
\begin{equation}
\varepsilon _{1}=i\sum_{j}\frac{\omega _{pj}^{2}}{\omega \omega _{cj}^{2}}%
\left( \nu _{jn}+\chi _{j\perp 2}\right) .
\end{equation}

Substituting expressions (31) and $\varepsilon _{1}$ from (21) in Equation
(18), we obtain in the case $\nu _{n}\gg \chi _{nz}$ a dispersion relation
which has a solution, 
\begin{equation}
\omega =\left[ k_{z}^{2}-\frac{\omega _{pi}^{2}}{c_{s}^{2}}\frac{\left(
v_{e0}-v_{i0}\right) ^{2}}{c^{2}}\right] ^{1/2}c_{A},
\end{equation}

\noindent where $c_{s}=\left[ 3\left( |q_{e}|T_{i}+q_{i}T_{e}\right)
/|q_{e}|m_{i}\right] ^{1/2}$ is the ion sound velocity. Using expressions
(31) and (32), we find in the case $\chi _{nz}\gg \nu _{n}$ the following
solution of the dispersion relation (18): 
\begin{equation}
\omega =i\left[ \sum_{j}\frac{\omega _{pj}^{2}}{\omega _{cj}^{2}}\left( \nu
_{jn}+\chi _{j\perp 2}\right) \right] ^{-1}\left[ \frac{\omega _{pi}^{2}}{%
c_{s}^{2}}\left( v_{e0}-v_{i0}\right) ^{2}-k_{z}^{2}c^{2}\right] .
\end{equation}

\noindent The perturbations (33) and (34) with wave numbers $k_{z}$ such
that $k_{z}<k_{th}$, where $k_{th}=\omega _{pi}\left| v_{e0}-v_{i0}\right|
/c_{s}c$,

\noindent are unstable due to different equilibrium velocities of electrons
and ions.

\section{On calculation of $E_{1rk}$}

To find the radial component of the electric field $E_{1rk}$ defined by
Equation (19), it is necessary to calculate the value $\varepsilon
_{3}\varepsilon _{\lambda z}+\varepsilon _{4}\varepsilon _{\delta z}$. The
terms $\varepsilon _{\delta z}$ and $\varepsilon _{\lambda z}$ have been
arisen due to perturbation of collisional frequencies proportional to the
number density perturbations. Using the expressions for these terms
containing in Equation (15) and the expressions for $\varepsilon _{3}$ and $%
\varepsilon _{4}$ from Equations (20), we obtain, 
\begin{eqnarray*}
\varepsilon _{3}\varepsilon _{\lambda z}+\varepsilon _{4}\varepsilon
_{\delta z} &=&i\frac{\omega _{pe}^{2}k_{z}^{2}n_{z}}{\omega ^{3}D_{z}^{2}}%
\frac{q_{i}}{m_{i}}\left( \frac{v_{i0}-v_{e0}}{c}\right) \sum_{j}\frac{%
\omega _{pj}^{2}}{\omega _{cj}}\frac{m_{j}}{q_{j}} \\
&&\times \left[ \left( \alpha _{iz}d_{e}-\nu _{en}\nu _{ni}d_{i}\right)
\beta _{jez}+\left( \alpha _{ez}d_{i}-\nu _{in}\nu _{ne}d_{e}\right) \beta
_{jiz}\right] ,
\end{eqnarray*}

\noindent where $\beta _{jlz}$, $l=e,i$, is defined in Section 4. We see
that the value $\varepsilon _{3}\varepsilon _{\lambda z}+\varepsilon
_{4}\varepsilon _{\delta z}\sim
(v_{e0}-v_{i0})(v_{e0}-v_{n0})+(v_{e0}-v_{i0})(v_{i0}-v_{n0})$. Thus,
polarization of perturbations also depends on the difference of equilibrium
velocities of species.

\section{On the resistivity in the standard MHD}

In the papers by Nekrasov (2007, 2008 a,b, and 2009 a,b) and in the present
paper, we study streaming instabilities of multicomponent rotating
magnetized objects, using the equations of motion and continuity for each
species. From Faraday's and Ampere's laws we obtain equations for the
electric field components (see Equation (15)). Such an approach allows us to
follow the movement of each species separately and obtain rigorous
conditions of consideration and physical consequences in specific cases
(see, e.g., Section (9)). This approach permits us to include various
species of ions and dust grains having different charges and masses. In some
cases, the standard methods used in the magnetohydrodynamics (MHD) leads to
conclusions that are different from those obtained by the method using the
electric field of perturbations. Indeed, let us consider the magnetic
induction equation. For simplicity, we take a two-component electron-ion
plasma embedded in the background magnetic field directed along the $z$
axis. The magnetic induction equation with the resistivity is obtained from
the momentum equation for the electrons at neglecting the electron inertia, 
\begin{equation}
\mathbf{0=-}\frac{e}{m_{e}}\mathbf{E-}\frac{e}{m_{e}c}\mathbf{v}_{e}\times 
\mathbf{B}-\nu _{ei}\left( \mathbf{v}_{e}-\mathbf{v}_{i}\right) ,
\end{equation}

\noindent where $\nu _{ei}$ ($\nu _{ie}$) is the electron-ion (ion-electron)
collisional frequency, and $-e$ is the electron charge. Replacing $\mathbf{v}%
_{i}-\mathbf{v}_{e}$ by the current $\mathbf{j}/en$ ($n_{e}=n_{i}=n$),
applying $\mathbf{\nabla \times }$ to Equation (35), and using Equations (4)
and (5), we obtain the well-known magnetic induction equation, 
\begin{equation}
\frac{\partial \mathbf{B}}{\partial t}=\mathbf{\nabla \times v}_{e}\times 
\mathbf{B+}\eta _{m}\mathbf{\nabla }^{2}\mathbf{B,}
\end{equation}

\noindent where $\eta _{m}=\nu _{ei}c^{2}/\omega _{pe}^{2}$ is the
coefficient of the magnetic diffusion or the resistivity. The momentum
equation for ions has the form, 
\begin{equation}
\frac{d\mathbf{v}_{i}}{dt}\mathbf{=}\frac{e}{m_{i}}\mathbf{E+}\frac{e}{m_{i}c%
}\mathbf{v}_{i}\times \mathbf{B}-\nu _{ie}\left( \mathbf{v}_{i}-\mathbf{v}%
_{e}\right) ,
\end{equation}

\noindent where $d/dt=\partial /\partial t+\mathbf{v}_{i}\cdot \mathbf{\nabla
v}_{i}$. It is seen from Equations (35) and (37) that if the electrons and
ions are magnetized, then $\mathbf{v}_{e\perp }\approx \mathbf{v}_{i\perp }$
(the sign $\perp $ denotes the transverse direction relatively to the
magnetic field). Usually, the electron velocity in Equation (36) is
substituted by the ion (or neutral) velocity. As a result, one solves the
standard two MHD equations, 
\begin{eqnarray}
\rho \frac{d\mathbf{v}}{dt} &=&\frac{1}{4\pi }\mathbf{\nabla \times B}\times 
\mathbf{B}, \\
\frac{\partial \mathbf{B}}{\partial t} &=&\mathbf{\nabla \times v}\times 
\mathbf{B+}\eta _{m}\mathbf{\nabla }^{2}\mathbf{B,}  \nonumber
\end{eqnarray}

\noindent where $\rho =m_{i}n$, $\mathbf{v=v}_{i}$.

From Equations (38) in the linear approximation, one obtains the dispersion
relations, 
\begin{equation}
\omega ^{2}+i\omega \eta _{m}k^{2}-k^{2}c_{A}^{2}=0,
\end{equation}

\noindent for magnetosonic waves, and 
\begin{equation}
\omega ^{2}+i\omega \eta _{m}k^{2}-k_{z}^{2}c_{A}^{2}=0,
\end{equation}

\noindent for Alfv\'{e}n waves. Here, $k^{2}=k_{\perp }^{2}+k_{z}^{2}$, $%
c_{A}=\left( B_{0}^{2}/4\pi \rho \right) ^{1/2}$ is the Alfv\'{e}n velocity.

Two consequences are followed from Equations (39) and (40):

1). Magnetosonic and Alfv\'{e}n waves are damped due to the resistivity.

2). The resistivity is isotropic since it is proportional to the total wave
number.

We derive now the dispersion relation, calculating the perturbed velocities
of electrons and ions through the electric field. For magnetized electrons
and ions, we find from Equations (35) and (37), 
\begin{eqnarray}
j_{1rk} &=&\frac{enc}{B_{0}}\frac{\omega }{\omega _{ci}}\left( -iE_{1rk}+%
\frac{\omega }{\omega _{ci}}E_{1\theta k}\right) , \\
j_{1\theta k} &=&\frac{enc}{B_{0}}\frac{\omega }{\omega _{ci}}\left(
-iE_{1\theta k}-\frac{\omega }{\omega _{ci}}E_{1rk}\right) ,  \nonumber \\
v_{e1z} &=&-\frac{e}{m_{e}\nu _{ei}}E_{1z},v_{i1z}=0.  \nonumber
\end{eqnarray}

\noindent We see that the collisional frequency $\nu _{ei}$ is absent in the
transverse current and determines only the electron velocity along the
background magnetic field. Using Equations (4) and (5) and neglecting the
small terms proportional to $\omega /\omega _{ci}$ in the round brackets of
Equations (41), we obtain the following dispersion relation: 
\begin{equation}
\left( \omega ^{2}-k^{2}c_{A}^{2}\right) \left( \omega ^{2}+i\omega \eta
_{m}k_{\perp }^{2}-k_{z}^{2}c_{A}^{2}\right) =0.
\end{equation}

\noindent We can conclude from Equation (42):

1). Magnetosonic waves are not damped because they have no field $E_{z}$.

2). Alfv\'{e}n waves are damped due to resistivity. The resistivity is
proportional to $k_{\perp }^{2}$ and is anisotropic. When $k_{\perp }=0$ Alfv%
\'{e}n waves also are not damped.

These results differ, in principle, from the one that follows from Equations
(39) and (40). From our viewpoint, Equation (42) takes into account the
physical mechanism of the collisional damping correctly.

\section{Discussion}

The neutrals participate in the electromagnetic perturbations only due to
collisions with the charged particles, electrons, ions, and dust grains. One
can say that the neutrals are as passive agent, ballast making difficult
perturbations of charged species. At the same time, the latter are only
active agents generating electromagnetic perturbations. Therefore, an
adequate description of multicomponent plasmas including the neutrals is to
express the neutral dynamics through the dynamics of charged species and
substitute induced velocities of neutrals into collisional terms of the
momentum equations for charged species. Then using Faraday's and Ampere's
laws, we can derive the dispersion relation in the linear approximation
and/or investigate nonlinear structures.

From expressions for induced velocities of neutrals, one can easily to find
rigorous conditions when the neutral dynamics is important or not, i.e.,
when there is strong (or sufficiently strong) or weak collisional coupling
of neutrals with charged species. In the last case, the neutrals are
immobile in the electromagnetic perturbations.

Let us write out conditions, which have been used for obtaining the
dispersion relation (18), and consider parameters of astrophysical objects,
for which these conditions can be satisfied. We will apply our results to
protostellar and protoplanetary disks. We have assumed that the electrons as
well as ions are magnetized, $\omega _{cj}\gg \nu _{jn}$. The condition $%
\omega _{ce}\left( >0\right) \gg \nu _{en}$ is, in general, satisfied in
astrophysical objects (Wardle \& Ng 1999). The rate coefficient for momentum
transfer by elastic scattering of electrons with neutrals is $<\sigma \nu
>_{en}=4.5\times 10^{-9}(T_{e}/30$ K$)^{1/2}$ cm$^{3}$ s$^{-1}$ (Draine et
al. 1983). As for ions, we should consider parameters to satisfy conditions $%
\omega _{ci}\gg \nu _{in}$ and $c_{Ai}^{2}\gg c_{si}^{2}\gg c_{A}^{2}$ or $%
c_{si}^{2}\gg c_{Ai}^{2}$ (the last two conditions see below). Thus, the
magnitude of the magnetic field must be in some limits. We take the standard
values $m_{i}=30m_{p}$ and $m_{n}=2.33m_{p}$ ($m_{p}$ is the proton mass).
The rate coefficient for momentum transfer $<\sigma \nu >_{in}$is equal to $%
<\sigma \nu >_{in}=1.9\times 10^{-9}$ cm$^{3}$ s$^{-1}$ (Draine et al.
1983). Then, for example, from conditions $\omega _{ci}\gg \nu _{in}$ and $%
c_{si}^{2}\gg c_{A}^{2}$ we obtain $\left( 3n_{n}T_{i}\right) ^{1/2}\gg
B_{0}\gg 0.43\times 10^{-12}n_{n}$ ($T_{i}$ is in the energetic units, $%
B_{0} $ is in G, and $n_{n}$ is in cm$^{-3}$), where we have used $%
q_{i}=-q_{e}$. Note that it is followed from the last inequalities that the
number density of neutrals is limited for a given ion temperature. If we
take $T_{i}$ $($K$)=700$ K, then we obtain $n_{n}\ll 1.57\times 10^{12}$ cm$%
^{-3}$. In this case, the neutral mass density $\rho _{n}=m_{n}n_{n}\ll
6.1\times 10^{-12}$ g cm$^{-3}$. This condition is applicable for early
stage of protoplanetary disks or in the surface layers, where the density is
lower and temperature is higher. In the dense inner parts of a disk, where $%
\rho _{n}\sim 10^{-10}-10^{-9}$ g cm$^{-3}$ (Hayashi et al. 1985; Wardle \&
Ng 1999), the ions can be unmagnetized and their viscosity will be of the
same form as that for neutrals. Our model is not applicable to such dense
regions.

For rotating protostellar cores of molecular clouds (protostellar disks), we
will adopt the following parameters: $n_{n}=10^{4}-10^{5}$ cm$^{-3}$, $%
n_{i}/n_{n}=10^{-5}-10^{-7}$ (e.g., Caselli et al. 1998; Ruffle et al. 1998;
Pudritz 2002), and $B_{0}=10$ $\mu $G (e.g., Goodman et al. 1993; Crutcher
et al. 1999; Caselli et al. 2002). For this magnetic field we obtain $\omega
_{ce}=1.76\times 10^{2}$ s$^{-1}$ and $\omega _{ci}=3.19\times 10^{-3}$ s$%
^{-1}$.

Now, we give some relationships which are useful at analysis of conditions
of consideration. For protoplanetary disks, we take $T_{e}=700$ K. Then we
obtain $\nu _{en}/\nu _{in}=158.73$, $\nu _{ne}/\nu _{en}=2.34\times
10^{-4}n_{e}/n_{n}$, $\nu _{ni}/\nu _{in}=12.88n_{i}/n_{n}$, $\nu _{ni}/\nu
_{ne}=3.47\times 10^{2}n_{i}/n_{e}$, $\nu _{en}/\nu _{n}=12.32n_{n}/n_{i}$, $%
m_{i}\nu _{in}/m_{e}\nu _{en}=3.47\times 10^{2}$, $\nu _{nn}/\nu _{in}=$ $%
6.94$ $\left( <\sigma \nu >_{nn}\sim <\sigma \nu >_{in}\right) $. For
protostellar disks, we take $T_{e}=70$ K. Then $\nu _{en}/\nu _{in}=50.19$, $%
\nu _{ni}/\nu _{ne}=1.1\times 10^{3}n_{i}/n_{e}$, $\nu _{en}/\nu
_{n}=3.9n_{n}/n_{i}$, and $m_{i}\nu _{in}/m_{e}\nu _{en}=1.1\times 10^{3}$.
We see that due to more low electron temperature, the role of
electron-neutral collisions in comparison to the ion-neutral collisions
becomes less in protostellar disks than that in protoplanetary disks. It can
be shown that inequalities (17) are wittingly satisfied under conditions
used in Section 6.

For small $k_{z}$ (Section 6.1), we will suppose that the ionization degree
is $n_{i}/n_{n}\ll 2.88(0.91)\times 10^{-3}$ for protoplanetary
(protostellar) disks (for the large $k_{z}$ (Section 6.2) this condition is
not necessary (see below)). In astrophysical objects, such as interstellar
medium, molecular clouds, and protostellar and protoplanetary disks the
ratio $n_{i}/n_{n}$ satisfies in general this condition. In this case, the
value $d_{e}=c_{sn}^{2}\nu _{en}$ (see inequality (20)). The value $d_{i}$
has the similar form, $d_{i}=c_{sn}^{2}\nu _{in}$, at $n_{i}/n_{n}\ll 1$. We
adopt that $T_{e}\sim T_{i}\sim T_{n}=T$. Then, for the thermal effects to
be important for all species, $k_{z}^{2}c_{sj,n}^{2}\gg \omega ^{2}$, the
condition $k_{z}^{2}c_{si}^{2}\gg \omega ^{2}$ will be sufficient. The
condition (20) for ions and electrons will be satisfied, if $\omega \nu
_{en}\gg k_{z}^{2}c_{se}^{2}$. The condition (24) takes the form $%
k_{z}^{2}c_{sn}^{2}\gg \omega \nu _{ni}$. In this case, there is weak
collisional coupling of neutrals with ions along the magnetic field (see
expressions (9)). It is followed from the last two inequalities that the
condition $m_{e}\nu _{en}\gg m_{n}\nu _{ni}$ must be satisfied for solutions
(26) and (27) to be realized. It was supposed above when we have obtained $%
d_{e}=c_{sn}^{2}\nu _{en}$. Under conditions at hand, the value $%
g=c_{sn}^{2}\nu _{in}\nu _{en}$ and $h^{2}=\omega _{pi}^{2}/\nu _{in}$ (see
Equation (25)). Then solution (26) takes the form, 
\begin{equation}
\gamma =\frac{\omega _{pi}^{2}}{\nu _{in}}\frac{\left( v_{e0}-v_{i0}\right)
^{2}}{c^{2}}.
\end{equation}%
Collecting all conditions given above, we find that solution (43) is
satisfied for wave numbers in the region%
\[
\frac{\gamma \nu _{en}}{c_{se}^{2}}\gg k_{z}^{2}\gg \max \left\{ \frac{%
\gamma ^{2}}{c_{A}^{2}},\frac{\gamma \nu _{ni}}{c_{sn}^{2}},\right\} .
\]%
Note that we consider the case in which $c_{si}^{2}\gg c_{A}^{2}$. Using
condition $\gamma /\nu _{ni}\ll 1$, we obtain from (43) that inequality $%
16.29\left\vert v_{e0}-v_{i0}\right\vert \ll \sqrt{n_{n}}$ ($v_{j0}$ is in
cm s$^{-1}$) must be satisfied. We see that for protostellar disks this
condition is not realized. For protoplanetary disks we take $n_{n}=5\times
10^{11}$ cm$^{-3}$. Then we obtain $\left\vert v_{e0}-v_{i0}\right\vert \ll
4.34\times 10^{4}$ cm s$^{-1}$. We further take $\left\vert
v_{e0}-v_{i0}\right\vert =3\times 10^{2}$ m s$^{-1}$, $n_{i}/n_{n}=10^{-9}$,
and $B_{0}=0.25$ G. In this case, $\omega _{pi}=5.37\times 10^{3}$ s$^{-1}$, 
$\omega _{ci}=79.75$ s$^{-1}$, $\nu _{in}=68.47$ s$^{-1}$, $\nu
_{ni}=8.82\times 10^{-7}$ s$^{-1}$, $\nu _{en}=1.09\times 10^{4}$ s$^{-1}$.
The Alfv\'{e}n velocity $c_{A}=5.05\times 10^{2}$ m s$^{-1}$, $c_{sn}=2.73$
km s$^{-1}$, $c_{si}=7.6\times 10^{2}$ m s$^{-1}$, and $c_{se}=1.78\times
10^{2}$ km s$^{-1}$. The growth rate $\gamma $ is equal to $\gamma
=4.2\times 10^{-7}$ s$^{-1}$. The wave number $k_{z}$ is in the limits, $%
3.8\times 10^{-7}$ m$^{-1}\gg k_{z}\gg 0.83\times 10^{-9}$ m$^{-1}$. The
wavelength of unstable perturbations $\lambda _{z}=2\pi /k_{z}$ has a range 
\[
7.57\times 10^{6}\textrm{ km}\gg \lambda _{z}\gg 1.65\times 10^{4} \textrm{km}
\]
Solution (27) has the form

\begin{equation}
\gamma =\nu _{ni}\left[ \frac{\omega _{ci}}{\nu _{in}}\frac{k_{z}\left(
v_{e0}-v_{i0}\right) }{\nu _{ni}}\right] ^{2/3}.
\end{equation}

\noindent This solution is satisfied for wave numbers in the region 
\[
\min \left\{ 1,\frac{\left( \nu _{en}\nu _{ni}\right) ^{3/4}}{\left(
k_{z}^{\ast }c_{se}\right) ^{3/2}},\left( \frac{\nu _{ni}}{k_{z}^{\ast }c_{A}%
}\right) ^{3}\right\} \gg \frac{k_{z}}{k_{z}^{\ast }}\gg \max \left\{ \left( 
\frac{\nu _{ni}}{k_{z}^{\ast }c_{si}}\right) ^{3},\left( \frac{\nu _{ni}}{%
k_{z}^{\ast }c_{sn}}\right) ^{3/2}\right\} , 
\]

\noindent where $k_{z}^{\ast }=\nu _{in}\nu _{ni}/\omega _{ci}\left\vert
v_{e0}-v_{i0}\right\vert $. From these inequalities we see, in particular,
that the condition $\left\vert v_{e0}-v_{i0}\right\vert /c_{si}<\nu
_{in}/\omega _{ci}$ must be satisfied. We can also conclude that $%
c_{si}^{2}\gg c_{A}^{2}$ in this case. The neutrals are strongly (weakly)
coupled with the ions across (along) the magnetic field in perturbations
(43) and (44). The viscosity of neutrals along (across) the magnetic field
is negligible in comparison to the thermal pressure ($\nu _{ni}$). The
viscosity of electrons and ions is negligible because $k_{z}^{2}\rho
_{j}^{2}\ll 1$, where $\rho _{j}=v_{Tj}/\omega _{cj}$ is the Larmor radius.

Let us consider solution (44) for specific parameters in protostellar and
protoplanetary disks. For protostellar disks we take $n_{n}=10^{5}$ cm$^{-3}$
and $n_{i}/n_{n}=10^{-7}$. Then we obtain $\nu _{in}=1.37\times 10^{-5}$ s$%
^{-1}$ and $\nu _{in}/\omega _{ci}=4.29\times 10^{-3}$. In this case, the
condition $\left\vert v_{e0}-v_{i0}\right\vert \ll c_{si}\nu _{in}/\omega
_{ci}$ takes the form $\left\vert v_{e0}-v_{i0}\right\vert \ll 1.03\times
10^{2}$ cm s$^{-1}$, where $c_{si}=2.4\times 10^{2}$ m s$^{-1}$. In real
situation, the value $\left\vert v_{e0}-v_{i0}\right\vert $ is most probably
larger and does not satisfied the last condition. Thus, solution (44) does
not realized in protostellar disks under conditions at hand.

Let us determine parameters of protoplanetary disks for which
solution (44) is satisfied. This solution exists if $\left\vert
v_{e0}-v_{i0}\right\vert /c_{si}<\nu _{in}/\omega _{ci}$. In the case when collisions of neutrals with ions do
not influence on the equilibrium velocity of neutrals $v_{n0}$, the
difference $\left\vert v_{e0}-v_{i0}\right\vert $ can be estimated as $%
\left\vert v_{e0}-v_{i0}\right\vert \approx v_{n0}\nu _{in}^{2}/\omega
_{ci}^{2}$ (Nekrasov 2009 b). Thus, the last inequality takes the form, $%
v_{n0}/c_{si}<\omega _{ci}/\nu _{in}$, where $c_{si}=7.59\times 10^{2}$ m s$%
^{-1}$ at $T_{i}=700$ K. We consider the region of rotating
(pre)protoplanetary disk where $v_{n0}\sim 10$ km s$^{-1}$. In this case, we
obtain condition $\nu _{in}/\omega _{ci}<7.59\times 10^{-2}$. We take $\nu
_{in}/\omega _{ci}=3\times 10^{-2}$ and $B_{0}=0.5\times 10^{-2}$ G. It is
followed from this that $\omega _{ci}=1.6$ s$^{-1}$, $\nu _{in}=4.8\times
10^{-2}$ s$^{-1}$, and $n_{n}=3.5\times 10^{8}$ cm$^{-3}$. The ionization
degree we take to be $n_{i}/n_{n}=10^{-8}$. Then $\nu _{ni}=6.17\times
10^{-9}$ s$^{-1}$. Other parameters are equal: $\omega _{pi}=4.49\times
10^{2}$ s$^{-1}$, $\left\vert v_{e0}-v_{i0}\right\vert =9$ m s$^{-1}$, $%
c_{A}=3.82\times 10^{2}$ m s$^{-1}$. Then we obtain $k_{z}^{\ast
}=2.05\times 10^{-11}$ m$^{-1}$. The wave number $k_{z}$ satisfies conditions%
\[
\left( \frac{\nu _{ni}}{k_{z}^{\ast }c_{A}}\right) ^{3}\gg \frac{k_{z}}{%
k_{z}^{\ast }}\gg \left( \frac{\nu _{ni}}{k_{z}^{\ast }c_{si}}\right) ^{3}. 
\]%
Under parameters at hand, we have $0.48\gg k_{z}/k_{z}^{\ast }\gg 0.06.$
Thus, the wavelength of perturbations is in the band 
\[
5.13\times 10^{9} \textrm{ km} \gg \lambda _{z}\gg 6.38\\textrm 10^{8}\textrm{ km.} 
\]%
For $k_{z}/k_{z}^{\ast }=0.4$ the growth rate (43) is equal to $\gamma
=0.54\nu _{ni}=3.33\times 10^{-9}$ s$^{-1}$.

Perturbations with more small $k_{z}$, $k_{z}^{2}c_{sn}^{2}\ll \omega \nu
_{ni}$ (see inequality (28)), have the growth rate (29). When additionally $%
k_{z}\ll s|v_{e0}-v_{i0}|/c$, where $s=\left( \omega _{pi}/c_{sn}\right)
\left( \nu _{ni}/\nu _{in}\right) ^{1/2}$, this growth rate is 
\begin{equation}
\gamma =\nu _{ni}\frac{\omega _{ci}}{\nu _{in}}\frac{\left\vert
v_{e0}-v_{i0}\right\vert }{c_{sn}}.
\end{equation}

\noindent The wave number $k_{z}$ is in the region 
\[
\min \left\{ \frac{\left( \gamma \nu _{ni}\right) ^{1/2}}{c_{sn}},s\frac{%
\left\vert v_{e0}-v_{i0}\right\vert }{c}\right\} \gg k_{z}\gg \frac{\gamma }{%
c_{si}}. 
\]

\noindent It is followed from here that conditions $\nu _{ni}\gg 12.88\gamma 
$ and $c_{si}^{2}\gg c_{A}^{2}$ must be satisfied. In these perturbations,
the neutrals are strongly coupled with ions both along and across the
magnetic field. The viscosity of species is also negligible in the case
(45). For parameters given above condition $\nu _{ni}\gg 12.88\gamma $
results in a large magnetization, $\nu _{in}/\omega _{ci}\lesssim 10^{-2}$,
for which the value $\left\vert v_{e0}-v_{i0}\right\vert $ is too small.
Thus, solution (45) is not available for protostellar and protoplanetary
disks.

Let us now consider the large $k_{z}$ when $k_{z}^{2}c_{si}^{2}\gg \omega
\nu _{in}$ (see inequality (30)). Under this condition the inequality $%
k_{z}^{2}c_{se}^{2}c_{sn}^{2}$ $\gg \omega d_{e}$ is satisfied at $%
n_{i}\lesssim n_{n}$. In this case, the neutrals are weakly coupled with
ions in the direction along the magnetic field (see Equations (9)).
Solutions considered below exist in media where $c_{si}^{2}\gg c_{Ai}^{2}$
or $12\pi n_{i}T_{i}\gg B_{0}^{2}$. This condition is not satisfied for
typical parameters in protostellar disks given above. However, near the
central star, an innermost part of the protoplanetary disk can be highly
ionized due to high temperature and have a high density. Note that a highly
ionized dense plasma disks exist around neutron stars, stellar black holes,
active galactic nuclei, and white dwarfs (see, e.g., Jin 1996 and references
therein). The last condition together with $\omega _{ci}\gg \nu _{in}$
results in $\left( 12\pi n_{i}T_{i}\right) ^{1/2}\gg B_{0}\gg 0.43\times
10^{-12}n_{n}$. We see from here that at $T\sim 10^{4}$ K and $n_{i}\sim
n_{n}$ the number density must be $n_{n}\ll 2.62\times 10^{14}$ cm$^{-3}$.
If we take $n_{n}\sim 10^{13}$ cm$^{-3}$, we obtain the range of the
magnetic field, $22.81$ G$\gg B_{0}\gg 4.3$ G. Let, at first, $\chi _{nz}\ll
\nu _{ni}$ or $k_{z}^{2}c_{sn}^{2}\ll \nu _{ni}\nu _{nn}$ when the neutral
viscosity does not play a role. Then the neutrals are strongly coupled with
ions across the magnetic field. Under conditions at hand we see that $\nu
_{ni}\gg 1.86\omega $ should be satisfied. In this case, the solution for $%
\omega $ has the form (33). For $k_{z}$ such that $k_{z}\ll k_{th}=\omega
_{pi}\left\vert v_{e0}-v_{i0}\right\vert /c_{s}c$, the growth rate is equal
to 
\begin{equation}
\gamma =\omega _{pi}\frac{c_{A}}{c_{s}}\frac{\left\vert
v_{e0}-v_{i0}\right\vert }{c}.
\end{equation}

\noindent The wave number $k_{z}$ is in the region 
\[
\min \left\{ k_{th},\frac{\left( \nu _{ni}\nu _{nn}\right) ^{1/2}}{c_{sn}}%
\right\} \gg k_{z}\gg \frac{\left( k_{th}c_{A}\nu _{in}\right) ^{1/2}}{c_{si}%
}. 
\]

Using parameters given above and taking $B_{0}=7$ G, we find: $\omega
_{pi}=7.59\times 10^{8}$ s$^{-1}$, $\omega _{ci}=2.23\times 10^{3}$ s$^{-1}$%
, $\nu _{in}=1.37\times 10^{3}$ s$^{-1}$, $\nu _{ni}=1.76\times 10^{4}$ s$%
^{-1}$, $c_{A}=3.16$ km s$^{-1}$, $c_{Ai}=0.88$ km s$^{-1}$, $c_{s}=4.06$ km
s$^{-1}$, $c_{si}=2.87$ km s$^{-1}$, and $c_{sn}=10.3$ km s$^{-1}$. From
condition $\nu _{ni}\gg 1.86\omega $ we can find the limit on the value $%
\left\vert v_{e0}-v_{i0}\right\vert \ll 4.81$ km s$^{-1}$. In the case $%
\left\vert v_{e0}-v_{i0}\right\vert =1$ km s$^{-1}$, we obtain $k_{th}=0.62$
m$^{-1}$.The growth rate (46) is equal to $\gamma =1.96\times 10^{3}$ s$%
^{-1} $. The wavelength of unstable perturbations has a range $1.1\times
10^{4}$ m$\gg \lambda _{z}\gg 10.06$ m. The viscosity of electrons and ions
is negligible since $k_{z}^{2}\rho _{j}^{2}\ll 1$.

The \noindent strong coupling of neutrals with ions in the transverse
direction is maintained at $\chi _{nz}\leq \nu _{ni}$. When $\chi _{nz}\gg
\nu _{ni}$ or $k_{z}^{2}c_{sn}^{2}\gg \nu _{ni}\nu _{nn}$ the neutral-ion
coupling is weak. In this case, the solution (34) at $k_{z}\ll k_{th}$ can
be written in the form 
\begin{equation}
\gamma =\frac{\omega _{ci}^{2}}{\nu _{in}\left( 1+1.2k_{z}^{2}\rho
_{i}^{2}\right) }\frac{\left( v_{e0}-v_{i0}\right) ^{2}}{c_{s}^{2}}.
\end{equation}

\noindent\ The condition $\chi _{nz}\gg \gamma $ is satisfied. The wave
number is in the region 
\[
k_{th}\gg k_{z}\gg \max \left\{ \frac{\left( \gamma \nu _{in}\right) ^{1/2}}{%
c_{si}},\frac{\left( \nu _{ni}\nu _{nn}\right) ^{1/2}}{c_{sn}}\right\} . 
\]

\noindent From these inequalities we see that $c_{si}^{2}\gg c_{Ai}^{2}$ $%
\left(\textrm{we assume }k_{z}^{2}\rho _{i}^{2}\leq 1\right)$. The growth
rate (46) can be larger than $\nu _{ni}$. At the same time, condition of
consideration is $\gamma \ll \nu _{in}$ (see Section (6)). Thus, the
condition $\left\vert v_{e0}-v_{i0}\right\vert /c_{s}\ll \nu _{in}/\omega
_{ci}$ must be satisfied. Note that solution (47) is the only case when the
viscosity of neutrals and ions is important. However, the neutrals are
weakly coupled with ions in this case and are immobile in perturbations.

\section{Conclusion}

In the present paper, we have studied electromagnetic streaming
instabilities\ in thermal viscous regions of rotating astrophysical objects,
such as, protostellar and protoplanetary magnetized accretion disks,
molecular clouds, their cores, and elephant trunks. However, the obtained
results can be applied to any regions of interstellar medium, where
different background velocities between electrons and ions can arise.

We have considered a weakly and highly ionized three-component plasma
consisting of electrons, ions, and neutrals. The cyclotron frequencies of
charged species have been supposed to be much larger than their collisional
frequencies with neutrals. The vertical perturbations along the background
magnetic field have been investigated. We have included the effect of
perturbation of collisional frequencies due to density perturbations of
species. We have shown that due to collisions of charged species with
neutrals and neutrals with charged species the latter experience the back
reaction on their perturbations. So far as the neutrals participate in the
electromagnetic perturbations only because of collisions with the charged
species, an adequate description of multicomponent plasmas including
neutrals is to express the neutral dynamics through the dynamics of charged
species and substitute induced velocities of neutrals into collisional terms
of the momentum equations for charged species. Then using Faraday's and
Ampere's laws, we can derive the dispersion relation and/or investigate
nonlinear structures.

The viscosity of magnetized species is important when $k_{z}^{2}\rho
_{j}^{2}\gtrsim 1$, where $\rho _{j}=v_{Tj}/\omega _{cj}$ is the Larmor
radius. The viscosity of neutrals is negligible in comparison to the thermal
pressure for the low frequency perturbations, $\omega \ll \nu _{nn}$. For
the one-dimensional perturbations along the magnetic field the thermal
pressure is present in the longitudinal perturbed velocities of species. In
the transverse velocity of neutrals, the viscosity of neutrals is important
when $\chi _{nz}\gg \nu _{ni}$ or $k_{z}^{2}c_{sn}^{2}\gg \nu _{ni}\nu _{nn}$%
. However, the neutrals are weakly coupled with ions in this case and are
immobile in perturbations.

The growth rates of perturbations in the wide region of the wave number have
been found. We have derived that the long wavelength part of spectrum can be
excited in medium, where $c_{Ai}\gg c_{si}\gg c_{A}$, and the short
wavelength perturbations are excited at $c_{si}\gg c_{Ai}$, where $c_{si}$, $%
c_{A}$, and $c_{Ai}$ are the ion thermal velocity, Alfv\'{e}n velocity
including the mass density of neutrals, and ion Alfv\'{e}n velocity
including the ion mass density, respectively. We have shown that the
viscosity plays the role only for the short wavelength edge of spectrum when
the neutrals have a weak collisional coupling with ions. In the cases of
strong neutral-ion coupling, the viscosity is negligible. The resistivity is
absent in the one-dimensional perturbations along the magnetic field.

On the simple example, we have demonstrated that there is discrepancies
between the standard MHD results and multicomponent approach at
consideration of the resistivity. From the latter approach it is followed
that the resistivity has an anisotropic nature. If the resistivity can play
an important role in damping of perturbations, this effect can results in
anisotropic turbulence.

Electromagnetic streaming instabilities considered in the present paper can
be a source of turbulence in thermal regions of astrophysical objects.

\bigskip

The insightful and constructive comments and suggestions of the anonymous
referee are gratefully acknowledged.

\paragraph{\noindent References\\}
\smallskip
\noindent \\
\noindent Balbus, S. A., \& Henri, P. 2008, ApJ, 674, 408

\noindent Braginskii, S. I. 1965, Rev. Plasma Phys., 1, 205

\noindent Braun, R., \& Kanekar, N. 2005, A\&A, 436, L53

\noindent Caselli, P., Benson, P. J., Myers, P. C., \& M. Tafalla, M. 2002, ApJ, 572,
238

\noindent Caselli, P.,\ Walmsley, C. M., Terzieva, R., \& Herbst, E. 1998,
ApJ, 499, 234

\noindent Crutcher, R. M., Roberts, D. A., Troland, T. H., \& Goss, W. M.
1999, ApJ, 515, 275

\noindent Draine, B. T., Roberge, W. G., and Dalgarno, A. 1983, ApJ, 264, 485

\noindent Goodman, A. A., Benson, P. J., Fuller, G. A., \& Myers, P. C. 1993, ApJ,
406, 528

\noindent Hayashi, C., Nakazawa, K., \& Nakagawa, Y. 1985, in Protostars and
Planets II, ed. D. C. Black \& M. S. Matthews (Tucson: Univ Arizona Press),
p. 1100

\noindent Jin, L. 1996, ApJ, 457, 798

\noindent Masada, Y., \& Sano, T. 2008, ApJ, 689, 1234

\noindent Nekrasov, A. K. 2007, Phys. Plasmas, 14, 062107

\noindent Nekrasov, A. K. 2008 a, Phys. Plasmas, 15\textbf{,} 032907

\noindent Nekrasov, A. K. 2008 b, Phys. Plasmas, 15, 102903

\noindent Nekrasov, A. K. 2009 a, Phys. Plasmas, 16, 032902

\noindent Nekrasov, A. K. 2009 b, ApJ, 695, 46

\noindent Pessah, M. E., \& Chan, C. 2008, ApJ, 684, 498

\noindent Pudritz, R. E. 2002, Science, 295, 68

\noindent Ruffle, D. P., Hartquist, T.W., Rawlings, J. M. C., \& Williams, D. A. 1998,
A\&A, 334, 678.

\noindent Stanimirovi\'{c}, S., \& Heiles, C. 2005, ApJ, 631, 371

\noindent Wardle, M., \& Ng, C. 1999, MNRAS, 303, 239

\noindent Yatou, H., \& Toh, S. 2009, Phys. Rev. E, 79, 036314

\bigskip

\end{document}